\documentstyle[twoside]{siamltex}
\newcommand{\fig}[3] %usage:\fig{file}{label}{caption}
{
 \begin{figure}[hbt]
 \begin{center}
 \input{#1}
 \end{center}
 \caption{#3}
 \label{#2}
 \end{figure}
}

\newenvironment{tabAlgorithm}[2]{
\setcounter{algorithmLine}{1}
\samepage
\begin{tabbing}
999\=\kill
#1 \ \ --- \ \ \parbox{4in}{\it #2}
}{
\end{tabbing}
}
\newcounter{algorithmLine}
\newcommand{\algline}{\\\thealgorithmLine\hfil\>\stepcounter{algorithmLine}}
\newcommand{\algcont}{\\\hfil\>}

\newcommand{\M}[1]{\mbox{#1}}
\newcommand{\Opt}[1]{\mbox{${\cal OPT}(#1)$}}
\newcommand{\MaxCycle}[1]{{\cal C}}

\newcommand{\ContractCycles}{\M{\sc Contract-Cycles}}
\newcommand{\Exchange}{\M{\sc Exchange}}

\newcommand{\Find}{\M{\sc Find}}
\newcommand{\MakeSet}{\M{\sc Make-Set}}
\newcommand{\Union}{\M{\sc Union}}

\newcommand{\NIL}{\M{\bf nil}}
\newcommand{\CURRENT}{\M{\bf current}}
\newcommand{\IF}{\M{\bf if }\=}
\newcommand{\THEN}{\M{\bf then }\=}
\newcommand{\ELSE}{\M{\bf else }}
\newcommand{\WHILE}{\M{\bf while }\=}
\newcommand{\FOR}{\M{\bf for }\=}
\newcommand{\DO}{\M{\bf do }\=}
\newcommand{\RETURN}{\M{\bf return }}

\newcommand{\ToActive}{\M{to-active}}
\newcommand{\ToRoot}{\M{to-root}}
\newcommand{\FromRoot}{\M{from-root}}
\newcommand{\DFS}{\M{\sc DFS}}
\newcommand{\ContractCycle}{\M{\sc Contract-Cycle}}

\newcommand{\Basicalg}{
\begin{tabAlgorithm}{$\ContractCycles_k(G)$}{}
\algline \FOR \= $i=k,k-1,k-2,...,2$
\algline        \> \WHILE \= the graph contains a cycle 
                                with at least $i$ edges
\algline        \>      \> Contract the edges on such a cycle.
\algline \RETURN the contracted edges
\end{tabAlgorithm}
}

\newcommand{\CMT}[1]{\={\em~~~--- #1 ---}}

\newcommand{\PracticalAlg}{
\begin{tabAlgorithm}{$\ContractCycles_3(G=(V,E))$}{Pseudo-code.}
\algline $S \leftarrow \{\}$
\algline Choose $r\in V$.
\algline $\DFS(r)$
\algline Add 2-cycles remaining in $G'/S$ to $S$.
\algline \RETURN $S$
\end{tabAlgorithm}

\begin{tabAlgorithm}{$\DFS(u)$}{ }
\algline $\ToActive[\Find(u)] \leftarrow \CURRENT$
\algline \FOR each vertex $w$ adjacent to $u$ \CMT{traverse edge $(u,w)$}
\algline   \> \IF ($w$ is not yet visited) \CMT{new vertex}
\algline   \>  \>  $\MakeSet(w)$ 
\algline   \>   \> $\ToActive[\Find(u)] \leftarrow 
                                \FromRoot[\Find(w)] \leftarrow (u,w)$
\algline   \>   \> $\DFS(w)$
\algline   \>   \> $\ToActive[\Find(u)] \leftarrow \CURRENT$
\algline   \> \ELSE \CMT{edge creates cycle in $G'/S$}
\algline   \>   \> \IF ($\Find(u) \ne \Find(w)$) 
                                        \CMT{cycle length at least 2}
\algline   \>   \>   \> $(x,y) \leftarrow \FromRoot[\Find(u)]$
\algline   \>   \>   \> \IF ($\Find(x) = \Find(w)$)
                                        \CMT{length two cycle through parent, $U-W-U$}
\algline   \>   \>   \>   \> $\ToRoot[\Find(u)] \leftarrow (u,w)$
                                        \CMT{record edge to parent}
\algline   \>   \>   \> \ELSE 
\algline   \>   \>   \>   \> $(x,y) \leftarrow \FromRoot[\Find(w)]$ 
\algline   \>   \>   \>   \> \IF ($\Find(x) \neq \Find(u)$)
                        \CMT{not a forward edge to child; therefore length of cycle $\ge 3$}
\algline   \>   \>   \>   \>   \> $\ContractCycle(w)$
\algline   \>   \>   \>   \>   \> $S \leftarrow S \cup \{(u,w)\}$
\algline $\ToActive[\Find(u)] \leftarrow \NIL$
\end{tabAlgorithm}

}

\newcommand{\cycleSubroutine}{
  \begin{tabAlgorithm}{$\ContractCycle(w)$}{ }
    \algline \WHILE ($\ToActive[\Find(w)] \neq \CURRENT$) \DO
    \algline \> \IF ($\ToActive[\Find(w)] = \NIL$) \THEN 
                          \CMT{Go up towards l.~c.~a.~along reverse edges.}
    \algline \> \> $(c,p) \leftarrow \ToRoot[\Find(w)]$
    \algline \> \> $a \leftarrow \ToActive[\Find(p)]$
    \algline \> \ELSE \CMT{Go down from l.~c.~a.~along active path.}
    \algline \> \> $(p,c) \leftarrow \ToActive[\Find(w)]$
    \algline \> \> $a \leftarrow \ToActive[\Find(c)]$
    \algcont \> \CMT{Contract parent $p$ and child $c$.}
    \algline \> $f \leftarrow \FromRoot[\Find(p)]$
    \algline \> $t \leftarrow \ToRoot[\Find(p)]$
    \algline \> $\Union(p,c)$
    \algline \> $\ToActive[\Find(w)] \leftarrow a$
    \algline \> $\FromRoot[\Find(w)] \leftarrow f$
    \algline \> $\ToRoot[\Find(w)] \leftarrow t$
  \end{tabAlgorithm}

}

\title{Approximating the Minimum Equivalent Digraph}
\author{Samir Khuller
\thanks{Computer Science Department and Institute for Advanced Computer
Studies, University of Maryland, College Park, MD~20742.
Research supported by NSF Research Initiation Award CCR-9307462.
E-mail~: {\tt samir@cs.umd.edu}.}
\and Balaji Raghavachari
\thanks{Computer Science Department, The University of Texas at
Dallas, Richardson, TX 75083-0688.
E-mail : {\tt rbk@utdallas.edu}.}
\and Neal Young
\thanks{
School of Operations Research and Industrial Engineering,
Cornell University, Ithaca, NY 14853-3801.
Part of this work was done while at UMIACS
and supported in part by NSF grants CCR-8906949 and CCR-9111348.
E-mail : {\tt ney@orie.cornell.edu}.}
}

\date{ }

\begin{document}

%\begin{titlepage}
\maketitle

\begin{abstract}
The MEG (minimum equivalent graph) problem is the following:
%samir: fixed this line
%``Given a strongly connected directed graph, 
``Given a directed graph, 
find a smallest subset of the edges
that maintains all reachability relations between nodes.''
This problem is NP-hard;
this paper gives an approximation algorithm
achieving a performance guarantee of about $1.64$ in polynomial time.
The algorithm achieves a performance guarantee of $1.75$
in the time required for transitive closure.

The heart of the MEG problem 
is the minimum SCSS (strongly connected spanning subgraph) problem
--- the MEG problem restricted to strongly connected digraphs.
For the minimum SCSS problem, 
the paper gives a practical, nearly linear-time implementation
achieving a performance guarantee of $1.75$.

The algorithm and its analysis 
are based on the simple idea of contracting long cycles.
The analysis applies directly to $2$-\Exchange, 
a general ``local improvement'' algorithm,
showing that its performance guarantee is $1.75$.
\end{abstract}



%\noindent      % next line automatically updated by RCS:
%{\tt $ $Source: /birdy/samir/submit/strong/RCS/journal.tex,v $ $, $ $Revision: 1.4 $ $}

\thispagestyle{empty}
%\end{titlepage}

\section{Introduction}
Connectivity is fundamental to the study of graphs and graph algorithms.
Recently, many approximation algorithms for
%samir: added the word ``minimum''
finding minimum subgraphs that meet given connectivity requirements
have been developed \cite{AKR,GSS,GW,KV,RK,WGMV}.
These results provide practical approximation algorithms for
NP-hard network-design problems via an  increased understanding of 
connectivity properties.  

Until now, the techniques developed
have been applicable only to {\em undirected} graphs. 
We consider a basic network-design problem
in {\em directed} graphs \cite{AGU,HNC,Hsu,MT} which is as follows:
given a digraph, find a smallest subset of the edges
(forming a {\em minimum equivalent graph (MEG)})
that maintains all reachability relations of the original graph.

When the MEG problem is restricted to strongly-connected graphs we call it
{\em the minimum SCSS (strongly connected spanning subgraph) problem}.
When the MEG problem is restricted to acyclic graphs we call it
{\em the acyclic MEG problem}. 
The MEG problem reduces in linear time \cite{CLR} to a single acyclic 
problem given by the so-called ``strong component graph'',
together with one minimum SCSS problem for each strong component
(given by the subgraph induced by that component).
Furthermore, approximating the MEG problem is linear-time equivalent 
to approximating both restricted versions.

\iffalse
We call the MEG problem restricted to strongly-connected graphs
{\em the minimum SCSS (strongly connected spanning subgraph) problem}.
Given a digraph,
the strong components can be found in linear time \cite{CLR}.
It is easily proved that the MEG problem thus reduces in linear time
to a single acyclic problem
(given by the so-called ``strong component graph'')
together with one minimum SCSS problem for each strong component
(given by the subgraph induced by that component).
Furthermore, approximating the MEG problem is linear-time equivalent 
to approximating both restricted versions.
\fi

Moyles and Thompson \cite{MT} observe this decomposition
and give exponential-time algorithms for the restricted problems.
Hsu~\cite{Hsu} gives a polynomial-time algorithm for the acyclic MEG problem.
%samir: RBK wanted to remove the next line
%and corrects some errors in the paper by Moyles and Thompson.

The related problem
of finding a {\em transitive reduction\/} of a digraph
--- a smallest set of edges yielding the same reachability relations
is studied by Aho, Garey and Ullman \cite{AGU}.
Transitive reduction differs from the MEG problem 
in that the edges in the transitive reduction are not required
to be in the original graph.
However, the transitive reduction problem decomposes 
just like the MEG problem into acyclic and strongly connected instances.
For any strongly connected instance, a transitive reduction is given 
by any Hamilton cycle through the vertices.
For an acyclic instance, the transitive reduction is unique
and, as Aho et al.~observe, is equivalent to the MEG problem:
it consists of those edges $(u,v)$
for which there is no alternate path from $u$ to $v$.
In fact, Aho, Garey and Ullman show that the transitive reduction
problem is {\em equivalent} to the transitive closure problem.
Thus, the acyclic MEG problem reduces to transitive closure.

The acyclic MEG problem can be solved in polynomial time,
whereas the minimum SCSS problem is NP-hard~\cite{GJ}.
Consequently, 
this paper focuses on approximation algorithms for the minimum SCSS problem.
By the observations of the preceding paragraphs,
the performance guarantees obtained for the minimum SCSS problem
carry over to the general MEG problem with the overhead of solving
a single instance of transitive closure.

\subsection{Our Results}
Given a strongly connected graph, our basic algorithm finds as long 
a cycle as it can, contracts the cycle, and recurses.
The contracted graph remains strongly connected.
When the graph finally collapses into a single vertex, 
the algorithm returns the set of edges
contracted during the course of the algorithm as the desired SCSS.

\iffalse
Intuitively, contracting a long cycle is good, 
because vertices are deleted from the graph at nearly
the same rate as edges are added to the solution.
Conversely, if no long cycle exists,
many edges are required to maintain strong connectivity.
\fi

The algorithm achieves a performance guarantee of any constant
%samir: rather than adding the +c stuff as the Ref asked for, I emphasized
% greater to make it clear that we can get any constant greater than...
{\em greater} than $\pi^2/6 \approx 1.645$ in polynomial time.
We give a nearly linear-time version 
that achieves a performance guarantee of $1.75$. 
We give examples showing lower bounds on the performance guarantees
of the algorithm.
For the general algorithm, the lower bounds are slightly above $1.5$.
For the nearly linear-time version, the lower bound is $1.75$, matching the
upper bound.

The performance guarantee analysis extends directly to
a simple ``local improvement'' algorithm called $2$-\Exchange.
$2$-\Exchange\ starts with the given digraph
and performs the following local improvement step 
as long as it is applicable:
find two edges in the current graph
that can be replaced by one edge from the original graph,
maintaining strong connectivity.
Similar local-improvement algorithms are natural candidates
for many optimization problems but often elude analysis.
%samir: the ref wants examples here. I know 2-exchange for TSP. Are there
% any other well-known examples...? RBK suggested low degree trees, max leaf trees
%N David Johnson essentially said the above in a discussion I had with him,
%N so I believe it even though I am ignorant of examples.
%N How about Simulated Annealing?  Not quite based on local improvement.
%N How about short augmenting paths for matching?  But it is analyzed.
We prove that the performance guarantee of $2$-\Exchange\ is $1.75$.

A natural improvement to the cycle-contraction algorithm
is to modify the algorithm to solve the problem optimally
once the contracted graph has no cycles longer than a given length $c$. 
For instance, for $c=3$, this modification
%samir
improves the performance guarantee to $\pi^2/6-1/36\approx 1.617$.
We use $\mbox{SCSS}_c$ to denote the minimum SCSS problem
restricted to digraphs with no cycle longer than $c$.
The minimum $\mbox{SCSS}_2$ problem is trivial.
%samir:
The minimum $\mbox{SCSS}_3$ problem can be solved in polynomial time, as 
shown by Khuller, Raghavachari and Young~\cite{KRY}.
%at least as hard as bipartite matching.
However, further improvement in this direction is limited:
we show that the minimum $\mbox{SCSS}_5$ problem is NP-hard.
In fact, we show that the minimum $\mbox{SCSS}_{17}$ problem is MAX SNP-hard.
This precludes the possibility of a polynomial-time approximation scheme, 
assuming P$\ne$NP~\cite{ALMSS}.

\subsection{Other Related Work}
The union of any incoming branching and any outgoing branching from 
the same root yields an SCSS with at most $2n-2$ edges (where $n$
is the number of vertices in the graph).
This is a special case of the algorithm given by Frederickson and 
J\'{a}J\'{a} \cite{FJ} that uses minimum weight branchings to achieve
a performance guarantee of 2 for weighted graphs. 
Since any SCSS has at least $n$ edges,
this yields a performance guarantee of $2$ for the SCSS problem.

Any {\em minimal\/} SCSS (one from which no edge can be deleted)
has at most $2n-2$ edges and also yields a performance guarantee of 2.
%samir: should we remove the next line since VLR claims that Simon's result
% is flawed. Why are we referencing all these papers anyway? should we check
% with simon?
%N Changed the line to avoid the problem.  Checking with Simon is a good idea.
\iffalse
A linear-time algorithm finding a minimal SCSS is given by Simon \cite{Simon}.
A parallel algorithm is given by Gibbons, Karp, Ramachandran,
Soroker and Tarjan~\cite{GKRST}.
\fi
 The problem of efficiently finding a minimal SCSS 
 is studied by Simon \cite{Simon}.
 Gibbons, Karp, Ramachandran, Soroker and Tarjan~\cite{GKRST}
 give a parallel algorithm.

A related problem in undirected graphs is to find a smallest
subset of the edges forming a biconnected
(respectively bridge-connected (i.e., 2-edge-connected)) spanning subgraph
of a given graph. These problems are NP-hard.
Khuller and Vishkin~\cite{KV} give a DFS-based algorithm
that achieves a factor of $\frac{5}{3}$ for biconnectivity and
$\frac{3}{2}$ for bridge-connectivity.
Garg, Santosh and Singla~\cite{GSS} subsequently improve
the approximation factors, using a similar approach, to $\frac{3}{2}$ 
and $\frac{5}{4}$, respectively.
None of these methods appear to extend to the minimum SCSS problem.

{\em Undirected} graphs having bounded cycle length have bounded tree width.
%rbk One reference for ``rich literature''? Sentence deleted.
%A rich literature \cite{ALS} exists concerning bounded-tree-width graphs.
Arnborg, Lagergren and Seese~\cite{ALS} have shown that
many NP-hard problems, 
including the minimum biconnected-spanning-subgraph problem,
have polynomial-time algorithms
when restricted to such graphs.

\section{Preliminaries}

To {\em contract} a pair of vertices $u,v$ of a digraph
is to replace $u$ and $v$
(and each occurrence of $u$ or $v$ in any edge)
by a single new vertex,
and to delete any subsequent self-loops and multi-edges.
Each edge in the resulting graph 
is identified with the corresponding edge in the original graph
or, in the case of multi-edges, the single remaining edge 
is identified with any one of the corresponding edges in the original graph.
%rbk Extra sentence added here for contracting an edge.
To contract an edge $(u,v)$ is to contract the pair of vertices $u$ and $v$.
To contract a set $S$ of pairs of vertices in a graph $G$
is to contract the pairs in $S$ in arbitrary order.
The contracted graph is denoted by $G/S$.
%rbk Another sentence for sets of edges.
Contracting an edge is also analogously extended to contracting a set of edges.

Let \Opt{G} be the minimum size of any subset of the edges 
that strongly connects $G$.
%Let $\MaxCycle{G}$ be the length of a longest cycle.
In general, the term ``cycle'' refers only to simple cycles.

\section{Lower Bounds on \Opt{G}}
We begin by showing that if a graph has no long cycles,
then the size of any SCSS is large.
\begin{lemma}[Cycle Lemma] \label{cycle-lemma}
For any directed graph $G$ with $n$ vertices, if a longest cycle
of $G$ has length $\MaxCycle{G}$, then
\[ \Opt{G} \ge \frac{\MaxCycle{G}}{\MaxCycle{G}-1}(n-1). \]
\end{lemma}

\begin{proof}
Starting with a minimum-size subset that strongly connects the graph,
repeatedly contract cycles in the subset until no cycles are left.
Observe that the maximum cycle length does not increase under contractions. 
Consequently, for each cycle contracted,
%samir: changed deleted to contracted....
the ratio of the number of edges contracted
to the decrease in the number of vertices
is at least $\frac{\MaxCycle{G}}{\MaxCycle{G}-1}$.
Since the total decrease in the number of vertices is $n-1$,
%rbk-2 changed deleted to contracted.
at least $\frac{\MaxCycle{G}}{\MaxCycle{G}-1}(n-1)$ edges are contracted.
\end{proof}

Note that the above lemma gives a lower bound which is existentially tight.
For all values of $\MaxCycle{G}$, there exist graphs for which the bound given
by the lemma is equal to \Opt{G}. Also note that $\MaxCycle{G}$ has a
trivial upper bound of $n$ and, using this, we get a lower bound of $n$ for
\Opt{G}, which is the known trivial lower bound.

\begin{lemma}[Contraction Lemma] \label{contraction-lemma}
For any directed graph $G$ and set of edges $S$,
\[ \Opt{G} \ge \Opt{G/S}.\]
\end{lemma}

\begin{proof}
Any SCSS of $G$, contracted around $S$ (treating the edges of $S$ as pairs), 
is an SCSS of $G/S$. 
\end{proof}

\section{Cycle-Contraction Algorithm}
The algorithm is the following.  Fix $k$ to be any positive integer.
\Basicalg

In Section~\ref{sec-imp}, we will show that the algorithm can be
implemented to run in $O(m\alpha(m,n))$ time for the case $k=3$
and in polynomial time for any fixed value of $k$.
It is clear that the edge set returned by the algorithm
strongly connects the graph.  The following theorem establishes an
upper bound on the number of edges returned by the algorithm.

\begin{theorem}
\label{mainthm}
$\ContractCycles_k(G)$ returns at most
$c_k \cdot \Opt{G}$ edges, where
%%% $c_k \cdot \Opt{G} - 1 + \frac{1}{k}$ edges, where
\[ {\pi^2\over 6} \le c_k 
        \le {\pi^2\over 6} + {1\over (k-1)k}.
\]
\end{theorem}
\begin{proof}
Initially, let the graph have $n$ vertices.
Let $n_i$ vertices remain in the contracted graph
after contracting cycles with $i$ or more edges ($i=k,k-1,...,2$).  

How many edges are returned?
In contracting cycles with at least $k$ edges,
at most $\frac{k}{k-1}(n-n_k)$ edges are contributed to the solution.
%rbk Added i<k
For $i<k$,
in contracting cycles with $i$ edges,
$\frac{i}{i-1}(n_{i+1}-n_i)$ edges are contributed.
The number of edges returned is thus at most
\[
\frac{k}{k-1}(n-n_k) + \sum_{i=2}^{k-1}\frac{i}{i-1}(n_{i+1}-n_i)
\le \left(1+\frac{1}{k-1}\right)n + \sum_{i=3}^k \frac{n_i-1}{(i-1)(i-2)}.
\]

Clearly $\Opt{G} \ge n$.  For $2\le i \le k$,
when $n_i$ vertices remain, no cycle has more than $i-1$ edges.
By Lemmas~\ref{cycle-lemma} and \ref{contraction-lemma},
$\Opt{G} \ge \frac{i-1}{i-2}(n_i-1)$.
Thus the number of edges returned, divided by \Opt{G}, is at most
\[
\frac{\left(1+\frac{1}{k-1}\right)n}{\Opt{G}}
+ \sum_{i=3}^{k} \frac{\frac{n_i-1}{(i-1)(i-2)}}{\Opt{G}}
\le
\frac{(1+\frac{1}{k-1})n}{n}
+ \sum_{i=3}^{k} \frac{\frac{n_i-1}{(i-1)(i-2)}}{\frac{i-1}{i-2}(n_i-1)}
=
\frac{1}{k-1} + \sum_{i=1}^{k-1} \frac{1}{i^2} = c_k.
\]

Using the identity (from \cite[p.75]{Knuth1}) 
$\sum_{i=1}^\infty \frac{1}{i^2} = \frac{\pi^2}{6}$, we get
%rbk-2 made the following an eqnarray
\begin{eqnarray*}
\frac{\pi^2}{6} &\le& c_k
=  \frac{\pi^2}{6} + \frac{1}{k-1} - \sum_{i=k}^\infty \frac{1}{i^2} \\
&\le& \frac{\pi^2}{6} + \frac{1}{k-1} 
        - \sum_{i=k}^\infty \frac{1}{i\,(i+1)} \\
&=& \frac{\pi^2}{6} + \frac{1}{k-1} - \frac{1}{k} \\
&=& \frac{\pi^2}{6} + \frac{1}{(k-1)k}.
\end{eqnarray*}
\end{proof}

If desired, standard techniques can yield more accurate estimates of $c_k$,
e.g.,
\(c_k = \frac{\pi^2}{6} + \frac{1}{2k^2} + O\left(\frac{1}{k^3}\right).\)
%N
If the graph initially has no cycle longer than $\ell$ ($\ell \ge k$),
then (as pointed out by an anonymous referee)
the analysis can be generalized to show a performance guarantee
of $\frac{k^{-1}-\ell^{-1}}{1-k^{-1}}+\sum_{i=1}^{k-1}1/^{i^2}$.
For instance, in a graph with no cycle longer than $5$,
the analysis bounds the performance guarantee (when $k=5$) by $1.464$.

Table~\ref{t-bounds} gives lower and upper bounds on the
performance guarantee of the algorithm for small values of $k$
and in the limit as $k\rightarrow\infty$.
The lower bounds are shown in the next subsection.

%rbk-2 added [h] below.
\begin{table}[h]
\begin{center}
\begin{tabular}{|c|c|c|c|}  \hline
 $k$    & Upper Bound   & Lower Bound \\ \hline
  3     & 1.750 & 1.750 \\
  4     & 1.694 & 1.666 \\
  5     & 1.674 & 1.625 \\
$\infty$& 1.645 & 1.500 \\
\hline
\end{tabular}
\end{center}
\caption{Bounds on the performance guarantee}
\label{t-bounds}
\end{table}

\subsection{Lower Bounds on the Performance Ratio}

In this section we present lower bounds 
on the performance ratio of $\ContractCycles_k(G)$.
The graph in Fig.~\ref{lbound} has $\frac{n}{2k-2}$ groups of vertices. 
Each group consists of a $(2k-2)$-cycle ``threaded'' with a $k$-cycle.

In the first iteration, 
$\ContractCycles_k(G)$ can contract the $k$-cycle within each group,
leaving the graph with only 2-cycles.
The algorithm subsequently must contract all the remaining edges. 
Thus, all the $(3k-2) \frac{n}{2k-2}-2$ edges are in the returned SCSS.
The graph contains a Hamilton cycle and the optimal solution is thus $n$. 
Hence, for arbitrarily large $n$, $1+ \frac{k}{2k-2} - 2/n$
is a lower bound on the performance guarantee of $\ContractCycles_k(G)$.
%rbk Extra line added below.
As $k$ approaches $\infty$, the lower bound tends to 1.5.

\fig{lbound}{lbound}{Bad example for $\ContractCycles_{k}(G)$.}

\newcommand{\LocalAlg}{
\begin{tabAlgorithm}{$k$-\Exchange$(G=(V,E))$}{Local improvement algorithm}
\algline $E' \leftarrow E$
\algline \WHILE \= the following improvement step is possible
\algline        \> Pick a set $E_k$ of $k$ edges in $E'$
                and a set $E_{k-1}$ of up to $k-1$ edges in $E$ 
\algcont        \> such that the set of edges $E'' = (E'-E_k)\cup E_{k-1}$
                forms an SCSS.
\algline        \> $E' \leftarrow E''$.
\algline \RETURN $E'$
\end{tabAlgorithm}
}

\section{$2$-Exchange\ Algorithm}
In this section, we use the cycle-contraction analysis
to show that $2$-\Exchange\ has a performance guarantee of $1.75$.
$2$-\Exchange\ is a special case of $k$-\Exchange,
which is defined as follows.

\LocalAlg

%rbk added "for fixed k" below; added "it" between "and" and "reduces".
Note that for fixed $k$, each step can be performed in polynomial time
and it reduces the size of $E'$, so $k$-\Exchange\ runs in polynomial time.
The following theorem shows that the
approximation factor achieved by $2$-\Exchange\ is $1.75$.

\begin{theorem}
The performance guarantee of $2$-\Exchange\ is $1.75$.
\end{theorem}
\begin{proof}
%samir: the ref wants some idea given before the proof.
% here is my attempt at giving the intuition (NEY may want to write
% it more crisply...
%N crisped...
We will show that the edges output by $2$-\Exchange$(G)$\ could be
output by $\ContractCycles_3(G)$.  
Thus, the performance guarantee of $1.75$ for $\ContractCycles_3$
carries over to $2$-\Exchange.
 
First we show that the performance guarantee is at most $1.75$.
Let $E'$ be the set of edges returned by $2$-\Exchange$(G=(V,E))$.
Run $\ContractCycles_3$ on the graph $G'=(V,E')$.
Let $H$ be the set of edges contracted during the first iteration
when cycles of at least three edges are contracted.
The resulting graph $G'/H$ is strongly connected and has only 2-cycles.
Such a graph has a tree-like structure. In particular,
an edge $(u,v)$ is present iff the reverse edge $(v,u)$ is present.

The important observation is that $G/H$ is equivalent to $G'/H$.
Clearly $G'/H$ is a subgraph of $G/H$; to prove the converse, 
suppose that some edge $(u,v)$ of $G/H$ was not in $G'/H$.
Consider adding edge $(u,v)$ to $G'/H$.
By the structure of $G'/H$,
$u$ and $v$ are not adjacent in $G'/H$
and for each edge on the path from $v$ to $u$
the reverse edge is also in $G'/H$.
If $(u,v)$ is added to $G'/H$,
these (at least two) reverse edges can be deleted from $G'/H$
without destroying the strong connectivity of $G'/H$.
Consequently, the original edge in $G$ corresponding to $(u,v)$
can be added to $G'$ and the original edges in $G'$
corresponding to the reverse edges can be deleted from $G'$
without destroying the strong connectivity of~$G'$.
This contradicts the fact that $E'$ was output by $2$-\Exchange$(G)$,
since $E'$ is eligible for an improvement step.

Next consider executing $\ContractCycles_3(G)$.
Since $G/H$ is equivalent to $G'/H$,
the sequence of cycles chosen in the first iteration of $\ContractCycles_3(G')$
could also be chosen by the first iteration of $\ContractCycles_3(G)$.
%The resulting graph $G/H$ would be equivalent to $G'/H$.
Similarly, the second iteration in $\ContractCycles_3(G')$
%rbk-2 : the following is shooting out of the page. What can we do? added \\
could be mimicked by \\ $\ContractCycles_3(G)$,
in which case $\ContractCycles_3(G)$ 
would return the same edge set as $\ContractCycles_3(G')$.
Since $E'$ is minimal (otherwise an improvement step applies),
the edge set returned is exactly $E'$.
Thus, the upper bound on the performance guarantee of $\ContractCycles_3$
from Theorem \ref{mainthm} is inherited by $2$-\Exchange.
\end{proof}

For the lower bound on the performance guarantee,
given the graph in Fig.~\ref{lbound2},
$2$-\Exchange\ can choose a number of edges
arbitrarily close to $1.75$ times the minimum.
%samir
There are $\frac{n}{4}$ groups with $4$ vertices in each group. First
observe that the graph has a directed Hamilton cycle.
The edges marked in Fig.~\ref{lbound2} form a solution that
$2$-\Exchange\ could terminate with. This solution clearly has
$\frac{7n}{4}$ edges. This gives the lower bound of $1.75$ on
the performance of the algorithm.

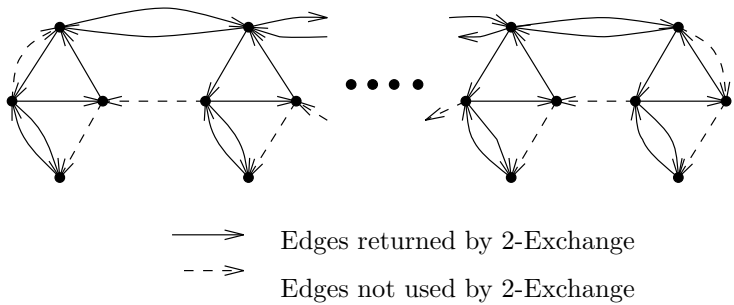
\fig{lbound2}{lbound2}{Worst-case example for $2$-\Exchange.}

\section{Implementation}
\label{sec-imp}

For any fixed $k$, 
$\ContractCycles_k$ can be implemented in polynomial time
using exhaustive search to find long cycles.
For instance, if a cycle of size at least $k$ exists,
one can be found in polynomial time as follows.
For each simple path $P$ of $k-1$ edges, 
check whether a path from the head of $P$ to the tail exists 
after $P$'s internal vertices are removed from the graph.
If $k$ is even, there are at most $m^{k/2}$ such paths;
if $k$ is odd, the number is at most $n\,m^{(k-1)/2}$.
It takes $O(m)$ time to decide if there is a path from
the head of $P$ to the tail of $P$. For the first iteration
of the for loop, we may have $O(n)$ iterations of the while loop.
Since the first iteration is the most time consuming,
the algorithm can be implemented in $O(n\,m^{1+k/2})$ time for even $k$ and 
$O(n^2\ m^{(k+1)/2})$ time for odd $k$.

\subsection{A practical implementation yielding $1.75$}
Next we give a practical, near linear-time implementation 
of $\ContractCycles_3$.
The performance guarantee achieved is $c_3 = 1.75$.
$\ContractCycles_3$ consists of two phases: 
(1) repeatedly finding and contracting cycles of three or more edges
(called {\em long} cycles),
until no such cycles exist,
and then (2) contracting the remaining 2-cycles.

\paragraph{High-level description of the algorithm.} 
To perform Phase (1),
the algorithm does a depth-first search (DFS) of the graph
from an arbitrary root.
During the search, the algorithm 
identifies edges for contraction by adding them to a set $S$. 
At any point in the search,
$G'$ denotes the subgraph of edges and vertices traversed so far.
The rule for adding edges to $S$ is as follows:
when a new edge is traversed,
if the new edge creates a long cycle in $G'/S$,
the algorithm adds the edges of the cycle to $S$.
The algorithm thus maintains that $G'/S$ has no long cycles.
When the DFS finishes, $G'/S$ has only 2-cycles.
The edges on these 2-cycles, together with $S$, are the desired SCSS.

Because $G'/S$ has no long cycles and the fact that the original graph
is strongly connected, $G'/S$ maintains a simple structure:

\begin{lemma}
\label{invariant lemma}
After the addition of any edge to $G'$ 
and the possible contraction of a cycle by adding it to $S$:
(i) The graph $G'/S$ consists of an outward branching
and some of its reverse edges.
(ii) The only reverse edges that might not be present
are those on the ``active'' path:
from the super-vertex containing the root to the super-vertex 
in $G'/S$ containing the current vertex of the DFS.
\end{lemma}

\begin{proof}
Clearly the invariant is initially true.
We show that each given step of the algorithm maintains the invariant.
In each case, if $u$ and $w$ denote vertices in the graph,
then let $U$ and $W$ denote the vertices in $G'/S$ containing $u$ and $w$, 
respectively.

%rbk Did not want the first part to be in bold.
\paragraph{{\it When the DFS traverses an edge $(u,w)$ to visit 
        a new vertex $w$:}}

Vertex $w$ and edge $(u,w)$ are added to $G'$.
Vertex $w$ becomes the current vertex.
In $G'/S$, the outward branching is extended
to the new vertex $W$ by the addition of edge $(U,W)$.
No other edge is added, and no cycle is created.
Thus, part (i) of the invariant is maintained.
The super-vertex containing the current vertex is now $W$,
and the new ``active path'' contains the old ``active path''.
Thus, part (ii) of the invariant is also maintained.

\fig{contract}{contract}{Contracted graph $G'/S$.}

\paragraph{{\it When the DFS traverses an edge $(u,w)$ and $w$ 
        is already visited:}}

If $U=W$ or the edge $(U,W)$ already exists in $G'/S$,
%samir: made changes ref asked for
then no cycle is created, $G'/S$ is unchanged,
and the invariant is clearly maintained.
Otherwise, the edge $(u,w)$ is added to $G'$
and a cycle with the simple structure 
illustrated in Fig.~\ref{contract} is created in $G'/S$.
The cycle consists of the edge $(U,W)$,
followed by the (possibly empty) path of reverse edges
%rbk added lca in parenthesis below and changed l. c. a. to lca.
from $W$ to the lowest-common-ancestor (lca) of $U$ and $W$,
followed by the (possibly empty) path of branching edges
%samir: W to U
from lca($U,W$) to $U$.
Addition of $(U,W)$ to $G'/S$ and contraction of this cycle
(in case it is a long cycle)
maintains part (i) of the invariant.
If the ``active path'' is changed, 
it is only because part of it is contracted,
so part (ii) of the invariant is maintained.

\paragraph{{\it When the DFS finishes visiting a vertex $w$:}}

No edge is added and no cycle is contracted, so part (i) is clearly maintained.
Let $u$ be the new current vertex, i.e., $w$'s parent in the DFS tree.
If $U=W$, then part (ii) is clearly maintained.
Otherwise, 
consider the set $D$ of descendants of $w$ in the DFS tree.
Since the original graph is strongly connected, 
some edge $(x,y)$ in the original graph
goes from the set $D$ to its complement $V-D$.
All vertices in $D$ have been visited,
so $(x,y)$ is in $G'$.
By part (i) of the invariant,
the vertex in $G'/S$ containing $x$ must be $W$, 
while the vertex in $G'/S$ containing $y$ must be $U$.
Otherwise the edge corresponding to $(x,y)$ 
in $G'/S$ would create a long cycle.
\end{proof}

The algorithm maintains the contracted graph $G'/S$ 
%rbk added citation to Tarjan's book here
using a union-find data structure \cite{Ta} to represent the vertices
in the standard way and using three data structures to maintain
the branching, the reverse edges discovered so far,
and the ``active path''.
When a cycle arises in $G'/S$,
it must be of the form described in the proof
of Lemma \ref{invariant lemma} and
illustrated in Fig.~\ref{contract}.
Using these data structures, 
the algorithm discovers it and, if it is long, contracts it
in a number of union-find operations proportional to the length of the cycle.
This yields an $O(m\alpha(m,n))$-time algorithm.

The vertices of $G'/S$ are represented in union-find sets as follows:
\begin{description}
\item[$\MakeSet(v)$:] 
        Adds the set $\{v\}$ corresponding to the new vertex of $G'/S$.
%rbk changed the definition of Find to match the other definitions.
% What was:
%\item[$\Find(v)$:] For any vertex $v$ in $G'$,
%       $\Find(v)$ returns the set corresponding to the vertex in $G'/S$ 
%       containing $v$.
\item[$\Find(v)$:] Returns the set in $G'/S$ that contains vertex $v$.
\item[$\Union(u,v)$:] Joins into a single set the two sets
corresponding to the vertices in $G'/S$ containing $G'$'s vertices $u$ and $v$.
\end{description}

The data structures representing the branching, reverse edges, and the
active paths, respectively are:
\begin{description}
%rbk How do we stop from-root, to-root and to-active from being in bold font?
\item[{\FromRoot[$W$]}:] For each branching edge $(U,W)$ in $G'/S$, 
        $\FromRoot[W] = (u,w)$ for some $(u,w)\in (U\times W) \cap E$.

\item[{\ToRoot[$U$]}:] For each reverse edge $(U,W)$ in $G'/S$,
        $\ToRoot[U] = (u,w)$ for some $(u,w)\in (U\times W)\cap E$.

\item[{\ToActive[$U$]}:] For each vertex $U$ on the ``active path'' in $G'/S$,
        $\ToActive[U] = (u,w)$ where $(u,w) \in (U\times W)\cap E$
        and $W$ is the child of $U$ for which the recursive DFS call
        is currently executing, unless no recursive DFS is executing,
        in which case $\ToActive[U] = \CURRENT$. 

        For all other vertices, $\ToActive[U] = \NIL$.
\end{description}

Pseudo-code for the algorithm is given in Figures~\ref{PractAlgFig}
and \ref{SubrFig}.
%rbk-2 made this into a table so that it appears on one page.
%\newpage
%N it's not a table now, though, did you really change it (?)
%N anyway it's moot, it got so long I split it into two figures

\begin{figure}[htb]
\PracticalAlg
\caption{Practical implementation of $\ContractCycles_3$}
\label{PractAlgFig}
\end{figure}

\begin{figure}[htb]
\cycleSubroutine
\caption{Subroutine $\ContractCycle$}
\label{SubrFig}
\end{figure}

%samir: Neal, the ref wants to say something about updating all the
% data-structures in the code. I think a couple of lines will suffice.
% what do you think?
%N added the details to the alg (please check them)

By the preceding discussion,
the algorithm implements $\ContractCycles_3$.
It is straightforward to show that it runs in $O(m\alpha(m,n))$ time.  
Hence, we have the following theorem.
\begin{theorem}
There is an $O(m\alpha(m,n))$-time approximation algorithm
for the minimum SCSS problem achieving a performance guarantee of $1.75$
on an $m$-edge, $n$-vertex graph.
\end{theorem}
Here $\alpha(m,n)$
%rbk-2 -- doesn't Ackermann have two n's?
is the inverse-Ackermann function 
associated 
%samir: fixed this
with the union-find data structure \cite{Ta}.

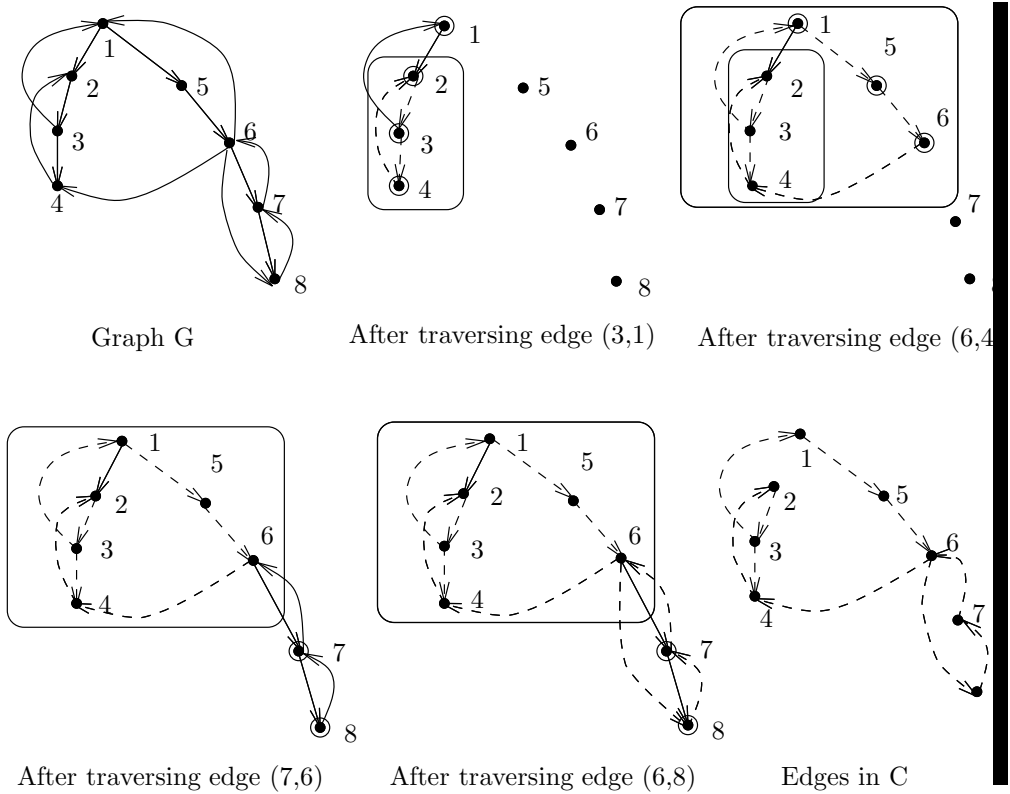
\fig{exam}{exam}{Example to illustrate execution of algorithm.}

\paragraph{Example to Illustrate Algorithm.}
In the example in Fig.~\ref{exam}, the algorithm begins the DFS from vertex 1. 
It visits vertices 2,3,4 and then traverses the reverse edge $(4,2)$. 
Since this edge creates a 3-cycle $(2,3), (3,4), (4,2)$ in $G'/S$
it contracts the cycle.
Next it traverses the reverse edge $(3,1)$, 
but does not contract it since it forms only a 2-cycle in the contracted graph.
Continuing the DFS, it visits vertices 5 and 6.
When it traverses the edge $(6,4)$, 
it discovers and contracts the cycle $(3,1),(1,5),(5,6),(6,4)$.
Next it visits vertices 7 and 8,
traversing the reverse edges $(8,7)$ and $(7,6)$. 
Traversing the edge $(6,8)$, it discovers and contracts
the 3-cycle $(8,7),(7,6),(6,8)$.  
In this example, no 2-cycles remain,
so it returns just the contracted edges.

\section{Potential Improvement of $\ContractCycles_k$}
A natural modification to $\ContractCycles_k$ would be to
stop when the contracted graph has no cycles of length more than some $c$
and somehow solve the remaining problem optimally.

For instance, for $c=3$, by following the proof of Theorem \ref{mainthm},
one can show that this would improve the performance guarantee
of $\ContractCycles_k$ to $c_k - 1/36$ (for $k \ge 4$),
matching the lower bound in Table \ref{t-bounds}.
(The lower bound given holds for the modified algorithm.)

%samir
This leads us to consider the minimum $\mbox{SCSS}_c$ problem
--- the minimum SCSS problem 
restricted to graphs with cycle length bounded by $c$.
The following theorem is shown in \cite{KRY}.

\begin{theorem}
There is a polynomial-time algorithm
for the $\mbox{SCSS}_3$ problem.
\end{theorem}

%samir: why do we need to tell the reader how hard the problem is?
% I think its enugh to say that it can be solved in P for this paper? comments?
%rbk-2 agrees; deleted the following sentence.
%It is not hard to show that the $\mbox{SCSS}_3$ problem
%is as hard as bipartite matching~\cite{KRY}.
%samir
%Also, one can assume w.l.o.g. that 
%rbk Rewrote the following line
%the given graph must have only 3-cycles,
%each of which has exactly two edges 
%{\em every} cycle in the graph has exactly three edges,
%and each cycle has exactly two edges 
%whose removal destroys strong connectivity.
%rbk Does the above line mean remove both edges or any one of them?

We make no conjecture concerning the $\mbox{SCCS}_4$ problem.
However, we next show that the $\mbox{SCCS}_5$ problem is NP-hard,
%samir
and that the $\mbox{SCSS}_{17}$ problem is MAX SNP-hard.

%rbk-2 created two new subsections.
\subsection{NP-hardness of SCSS$_5$}
We prove the following theorem.

\begin{theorem}
The minimum $\mbox{SCSS}_5$ problem is NP-hard.
\end{theorem}

\begin{proof}
%samir (general changes)
The proof is by a reduction from SAT \cite{GJ}.
Fix an arbitrary CNF formula $F$.
We will build a rooted digraph 
such that any SCSS contains all the edges out of the root ($d$ of them)
and such that $F$ is satisfiable iff there exists an SCSS $E'$
in which each of the remaining
$n-1$ non-root vertices has out-degree equal to one. Thus the formula will
be satisfiable if and only if there is an SCSS with $n-1+d$ edges.

%rbk-2 -- Why do we need the picture in the right?  Does the caption
%under it make complete sense?  There is some overlap between the
%right caption and the arrow to its left.
\fig{np}{NP}{Variable gadget for NP-hardness proof.}

The graph has a fixed root vertex $r$
and a vertex for each clause in $F$ (these vertices are not shown
in Fig..~\ref{NP}).
Each clause vertex has a return edge to the root.
For each variable in $F$, 
the graph has an instance of the gadget illustrated in Fig.~\ref{NP}.
The edges into the gadget come from the root.
Each such edge is present in any SCSS.
The edges out of the gadget are alternately labeled $+$ and $-$.
For every clause with a positive instance of the variable,
one of the $+$ edges goes to the clause vertex.
For every clause with a negative instance of the variable,
one of the $-$ edges goes to the clause vertex.
Unassigned $+$ and $-$ edges go to the root.
(The gadget is easily enlarged to allow any number of occurrences.)

The key property of the gadget is that if every (non-root) vertex
has out-degree one in some SCSS, 
%rbk-2 changed anti- to counter-
then either all of the counter-clockwise edges are in the SCSS
(corresponding to the variable being true)
or all of the clockwise edges are in the SCSS
(corresponding to the variable being false).
Thus, given any SCSS of $d+n-1$ edges, 
%rbk-2 added the following as per referee's comments
where $d$ is the outdegree of the root and
$n$ is the number of vertices in the digraph constructed,
it is easy to construct a satisfying assignment for $F$.
Conversely, given any satisfying assignment for $F$,
is is easy to construct an SCSS of size $d+n-1$.
\end{proof}

%\newpage

\newcommand{\maxlen}{{17}}
%rbk-2 added new subsection heading.
\subsection{MAX SNP-hardness of SCSS$_\maxlen$}
Next we consider the MAX SNP-hardness of the problem.
%rbk-2 added the following as guidelines for the reader.
The proof uses a reduction from the vertex-cover problem
in bounded-degree graphs to the SCSS problem.
Since the proof follows closely the reduction from vertex cover to
Hamiltonian circuits (see \cite{GJ}), it is suggested that the reader
study this reduction before reading this subsection.
It is known that the problem of finding a minimum vertex cover is MAX SNP-hard 
in graphs whose maximum degree is bounded by seven \cite{PY}.

Let $G$ be a connected, undirected graph
whose maximum degree is bounded by seven.
Let $G$ have $m$ edges and $n$ vertices.
We construct a digraph $D$ with $2m+1$ vertices
and no cycle longer than \maxlen.
Any vertex cover of $G$ of size $s$ will yield 
an SCSS of $D$ of size $2m+s$, and vice versa.
%N
\iffalse
Since the degree of $G$ is bounded, $m = O(n) = O(s)$ 
and it is easily verified that this yields
an L-reduction from degree-bounded vertex cover 
to the minimum $\mbox{SCSS}_\maxlen$ problem.
\fi
We then show that, since $G$ has $O(n)$ edges,
this yields an L-reduction (i.e., an approximation-preserving
reduction \cite{PY}).

\subsubsection{The construction of $D$}

Applying Vizing's theorem \cite{Vizing},
color the edges of $G$ in polynomial time
with at most eight colors so that no two edges
incident to a vertex share the same color.
Let the colors of the edges be one of $\{1,2,...,8\}$.

%rbk-2 -- the following paragraph is new.
The construction begins with a special ``root'' vertex $r$ in $D$
with outgoing arcs to each of $n$ vertices $v_1,\ldots,v_n$,
corresponding to the vertices of $G$.
An arc from $r$ to $v_i$ is labeled $v_i^+$.
Each of these vertices form the
beginning of a path corresponding to each vertex of $G$.

\fig{cover-test}{cover}{A cover-testing component.}

As the construction proceeds, 
each vertex in $G$ will have a ``current vertex'' in $D$,
initially the start vertex.
%rbk-2 changed increasing to non-decreasing
For each edge $(u,v)$, in order of non-decreasing color,
add a ``cover-testing gadget'' to $D$,
as illustrated in Fig.~\ref{cover}.
Specifically, add two new vertices $x$ and $y$.
Add two edges into $x$:
the first, labeled $u^+$, from the current vertex of $u$;
the second, labeled $u^-$, from $y$.
Similarly, add two edges into $y$:
the first, labeled $v^+$, from the current vertex of $v$;
the second, labeled $v^-$, from $x$.
Make $y$ the new current vertex of $u$;
make $x$ the new current vertex of $v$.
Finally, after all edges of $G$ have been considered,
for each vertex $v$ in $G$, 
add an edge labeled $v^+$ from its final current vertex to the root.
The gadgets are implicitly layered, 
with each gadget being assigned to a layer 
corresponding to the color of the associated edge in $G$.

\begin{lemma} \label{cycle-length-lemma}
The graph $D$ constructed above has no cycle with more than \maxlen\ edges.
\end{lemma}

\begin{proof}
We first assign numbers to the vertices of $D$.
The root $r$ is assigned the number 0.
The construction above proceeds in the order of increasing color
of the edges of $G$.
When considering an edge $(u,v)$ of color $c$, we add two new vertices:
$x$ is added to $v$'s path and $y$ is added to $u$'s path.
We assign the vertices $x$ and $y$ the number $c$.
Consider any cycle $X$ of length greater than two in $D$.
It is clear that such a cycle must pass through $r$,
since $D$ is layered.
Hence the cycle is of the form $(r, x_1, x_2, \ldots, x_k, r)$.
Because we considered the edges in order of increasing color,
the numbers assigned to the vertices in $X$ increases at least
every two steps in any path in $D$ (not including $r$).
In other words, the numbers assigned to the vertices $x_1,\ldots,x_k$
forms a non-decreasing sequence in which no three consecutive vertices
get the same number.  Since the edges of $G$ were colored with 8 colors,
the numbers assigned to the vertices of $D$ range from 0 to 8 (only $r$ gets
the number 0).  Combining all these, the length of the cycle $X$ is at most 17.
\end{proof}

\subsubsection{The analysis}
We now show that every vertex cover of $G$ has a corresponding SCSS in $D$.
The proof is similar to the corresponding proof (in the reduction from vertex
cover to Hamiltonian circuits) that every vertex cover has a corresponding
Hamiltonian circuit.
Consider an arbitrary vertex cover $S$ of $G$.
The idea is to choose in the SCSS the paths corresponding to $S$ in $D$.
The paths of the vertices of $V-S$ are yet to be connected.
Since $S$ forms a vertex cover, the vertices in the paths of $V-S$
can be connected using the cover-testing components.

\begin{lemma} \label{vc-to-scss-lemma}
Given a vertex cover of size $s$ of $G$,
an SCSS of $D$ of size $2m+s$ can be constructed.
\end{lemma}

\begin{proof}
Construct a subgraph $H$ of $D$ as follows.
For each vertex $u$ in $G$,
let $d$ be the degree of $u$ in $G$.
If $u$ is in the vertex cover,
add the $d+1$ edges labeled $u^+$ in $D$ to $H$.
Otherwise, add the $d$ edges labeled $u^-$ in $D$ to $H$.
It is easy to verify that $H$ has the following properties:
\begin{enumerate}
\item $H$ has $2m+s$ edges.
\item $H$ has no cycles of length 2.
\item Every vertex of $H$ has at least one outgoing
        and at least one incoming edge.
\end{enumerate}
As mentioned earlier, $D$ is layered and every cycle of length greater
than 2 contains $r$. Therefore Property~2 above implies that
every cycle of $H$ passes through $r$.
By the above conditions, $H$ contains a path from $r$ to every vertex $v$
and another path from $v$ to $r$, and is therefore strongly connected.
To obtain a path from any $v$ to $r$, start from $v$ and keep traversing
an outgoing edge (which exists by Property~3) from the current vertex.
Such a path must eventually reach $r$ because $r$ is contained
in every cycle of $H$.
Hence $H$ satisfies the lemma.
\end{proof}

We now show that every SCSS of $D$ corresponds
to a vertex cover of $G$.
The proof works by showing that any SCSS can be
converted into a ``canonical'' SCSS whose size is no larger,
that corresponds to a vertex cover of $G$.

\begin{lemma} \label{scss-to-vc-lemma}
Given an SCSS in $D$ of size $2m+s$,
a vertex cover of $G$ of size $s$ can be constructed.
\end{lemma}

\begin{proof}
First, as long as some non-root vertex $y$ has
both of its incoming edges in the SCSS, modify the SCSS as follows. 
Let $(x,y)$ be the edge labeled $v^-$ for some $v$.
Remove the edge $(x,y)$ and add the other edge out of $x$,
if it is not already present.
Alternatively, if some non-root vertex $x$ has
both of its outgoing edges in the SCSS, 
remove the edge $(x,y)$ and add the other edge into $y$.
Repeat either modification as long as applicable.

By the layering of $D$,
each modification maintains the strong connectivity of the SCSS.
Clearly none of the modifications increases the size. 
Each step reduces the number of edges labeled $u^-$ for some $u$
in the SCSS, so after at most $2m$ steps, 
neither modification applies, and in the resulting SCSS
every non-root vertex has exactly one incoming edge
and one outgoing edge in the SCSS.

An easy induction on the layering shows that for any vertex $v$ in $G$,
either all of the edges labeled $v^+$ in $D$ are in the SCSS or none are,
in which case all of the edges labeled $v^-$ are in the SCSS.
Let $C$ be the set of vertices in $G$ of the former kind.
It is easy to show that the size of the SCSS is $2m+|C|$, so that $|C| \le s$.
For every edge $(u,v)$ in $G$,
the form of the gadget ensures that 
at least one of the two endpoints is in $C$.
Hence, $C$ is the desired cover.
\end{proof}

\begin{theorem}
The minimum $\mbox{SCSS}_\maxlen$ problem is MAX SNP-hard.
\end{theorem}

\begin{proof}
Let $G$ be an arbitrary undirected graph $G$
whose maximum degree is bounded by seven.
Let $G$ have $m$ edges and $n$ vertices.
Construct the digraph $D$ as shown earlier.
By Lemma~\ref{cycle-length-lemma}, $D$ has no
cycles greater than \maxlen.
By Lemma~\ref{vc-to-scss-lemma},
any vertex cover of $G$ of size $s$ can be used to obtain
an SCSS of $D$ of size $2m+s$.
Conversely, by Lemma~\ref{scss-to-vc-lemma},
an SCSS of $D$ of size $2m+s$ can be used to obtain
a vertex cover of $G$ of size $s$.
Since the degree of $G$ is bounded, $m = O(n) = O(s)$ 
and it is easily verified that this yields
an L-reduction from degree-bounded vertex cover 
to the minimum $\mbox{SCSS}_\maxlen$ problem.
\end{proof}

\section{Open Problems}
An obvious problem is to further characterize 
the various complexities of the minimum $\mbox{SCSS}_k$ problems.

The most interesting open problem is to obtain a
performance guarantee that is less than 2
for the weighted strong connectivity problem (as mentioned earlier,
the performance factor of 2 is due to Frederickson and J\'{a}J\'{a} \cite{FJ}).
%rbk Too informal; changed sentence.
%We suspect that such an algorithm would have implications
Such an algorithm may have implications
for the weighted 2-connectivity problem \cite{KV} in undirected graphs as well.

The performance guarantee of $k$-\Exchange\ 
probably improves as $k$ increases.
Proving this would be interesting ---
similar ``local improvement'' algorithms
are applicable to a wide variety of problems.

%rbk-2 changed the citations to SIAM standard.

\end{document}